\documentclass[aps,twocolumn,groupedaddress]{revtex4}
\usepackage{epsfig}
\usepackage{graphicx}
\usepackage[T1]{fontenc}
\usepackage{ae}
\usepackage{color}
\usepackage[latin1]{inputenc}
\usepackage{amssymb,amsbsy,amsmath}
\usepackage{bbm}
\begin{document}

%%%%%%%%%%%%%%%%%%%%%%%%%%%%%%%%%%%%%%%%%%%%%%%%%%%%%%%%%%%%%%%%%%%%%%%%%%%%%

\title[Snellius meets Schwarzschild]
{Snellius meets Schwarzschild -- Refraction of brachistochrones and time-like geodesics}

\author{Heinz-J\"urgen Schmidt$^1$
}
\address{$^1$  Universit\"at Osnabr\"uck,
Fachbereich Physik,
 D - 49069 Osnabr\"uck, Germany}

%\tableofcontents

\begin{abstract}
The brachistochrone problem can be solved either by variational calculus or by a skillful application of the Snellius' law of refraction.
This suggests the question whether also other variational problems can be solved by an analogue of the refraction law.
In this paper we investigate the physically interesting case of free fall in General Relativity that can be formulated as a variational problem
w.~r.~t.~proper time. We state and discuss the corresponding refraction law for a special class of spacetime metrics including the Schwarzschild metric.
\end{abstract}

\maketitle

%%%%%%%%%%%%%%%%%%%%%%%%%%%%%%%%%%%%%%%%%%%%%%%%%%%%%%%%%%%%%%%%%%%%%%%%%%%%%%%%%%%%%%%%%%%%%%%%%%%%%%%%%%%%%%%%%%%%%%%%%%%%%%%
\section{Introduction}\label{sec:I}
%%%%%%%%%%%%%%%%%%%%%%%%%%%%%%%%%%%%%%%%%%%%%%%%%%%%%%%%%%%%%%%%%%%%%%%%%%%%%%%%%%%%%%%%%%%%%%%%%%%%%%%%%%%%%%%%%%%%%%%%%%%%%%%

Sometimes the same result can be derived in different ways. In physics we may benefit from considering side-by-side approaches
solving the same problem since they may have varying virtues; one approach may be more general and the other may provide more physical insight.

For example, the brachistochrone problem posed by Johann Bernoulli in 1696 often appears in textbooks as an exercise in
the variational calculus and is solved via the corresponding Euler-Lagrange equation.
However, at that time the variational calculus was not yet developed (the corresponding textbook of Leonhard Euler \cite{E44} was only published in 1744).
Hence it is not surprising that Bernoulli solved the problem in a different way: He considered the analogy with a beam of light in a medium
with varying refractive index that, according to Fermat's principle (stated in 1662), chooses the path between two points that takes the least time.
At first sight this only means that we have two different physical variational problems that have the same solution, but for the
optical problem there exists a local law determining the path of light: the Snellius' law of refraction (1632), formulated in a way suitable
for a smoothly varying refractive index. Bernoulli succeeded in showing that the cycloid curve satisfies the refraction law for the problem under consideration
and thus solved the brachistochrone problem.

Undoubtedly, the variational approach is straightforward and applicable to a large class of similar problems. Given the ubiquity
of variational principles in modern physics it is in order that students have to learn how to find and (sometimes) solve the Lagrange equations
and the brachistochrone problem could serve as an entertaining exercise. However, Bernoulli's solution has a unique appeal in so far as it uses the
transfer from other branches of physics and solves a problem not by setting into operation a heavy machinery but by a clever application of a simple law.
Our intuition is not well-trained to solve such kind of variational problems. One may concede that the naive choice of a straight line between two
given points $A$ and $B$ does not yield the path of least time and that it is a promising strategy to speed up early even if the path is thereby prolonged.
But the detailed form of the optimal curve is not intuitively clear. Recall that Galileo Galilei also considered the brachistochrone problem and conjectured
its solution to be the circle. On the other hand, the Snellius' law is intuitively clear in the light of Fermat's principle, at least qualitatively.
Hence, if we divide the space between $A$ and $B$ into a large number of layers with different but constant refractive indices
and imagine the refraction of light at each boundary between the layers,
the form of the optimal curve will become plausible, including the possible minimum of the brachistochrone due to total internal reflection of the light beam.

In this paper we will try to transfer Bernoulli's idea to another problem in physics that can be cast into the form of a variational problem: The motion of
a freely falling point particle according to General Relativity.
Given two points $A$ and $B$ in spacetime the actual world-line taken by the particle is the time-like geodesic
that assumes the maximal proper time $\tau$ among all time-like curves joining $A$ and $B$. Variational calculus yields the differential equation governing the
freely falling particle, namely the geodesic equation
\begin{equation}\label{I1}
 \frac{d^2 x^\lambda}{d\tau^2}+\Gamma_{\mu\nu}^\lambda\,\frac{d x^\mu}{d\tau}\,\frac{d x^\nu}{d\tau}=0
 \;,
\end{equation}
see, e.~g.~, \cite{W84} (4.4.18). In the limit of low velocities (compared with the speed of light) and linearly varying gravitational
potential the solutions of (\ref{I1}) would include the parabolic trajectories well-known from high school physics. However, due to the complicated
way to derive these elementary results one could hardly justify the claim that the parabolic trajectory is intuitively understood in terms of General Relativity.
In the light of Bernoulli's problem one would have to look for an analogy of the Snellius' law for time-like geodesics. This would require
a situation where the spacetime metric $g_{\mu\nu}$ is constant within two regions ${\mathcal A}$ and ${\mathcal B}$ separated by a space-like surface. If
the metric in ${\mathcal A}$ and ${\mathcal B}$  would be the same we have essentially a situation described by Special Relativity.
Then the straight time-like line joining two given points
$A\in{\mathcal A}$ and $B\in{\mathcal B}$ would realize the maximal proper time among all time-like curves joining $A$ and $B$.
But if the spacetime metric in ${ \mathcal A}$ and ${\mathcal B}$ is different then the maximal proper time is realized by a world line
composed of two straight lines within ${ \mathcal A}$ and ${\mathcal B}$ with different velocities. The equation relating these velocities
to the change of the metric would be the "Snellius' law for time-like geodesics" we are looking for.
Intuitively, the particle wants to spend more time in the region with higher gravitational potential, say, in ${\mathcal B}$, since there its internal clock runs faster and it can gain more proper time than in the region with lower potential, say, ${\mathcal A}$. But it would not be a good idea to rush
too quickly into ${\mathcal B}$ since this would slow down the internal clock according the time dilation effect already known from Special Relativity.
This qualitative explanation of free fall according to General Relativity has also be given in \cite{R17}.

Admittedly, the situation of two regions ${ \mathcal A}$ and ${\mathcal B}$ with a jump of the metric at its boundary is un-physical and only a fictitious situation in which we can argue in a simplified way. This is so because a physical metric has to satisfy Einstein's field equations and the described jump could only be realized
by a double layer of positive and negative mass density and therefore has to be ruled out. This is in contrast to the situation in geometrical optics
where two regions with different but constant refractive indices can be easily realized.
Nevertheless, the sequence of different finer and fine unphysical layers has a physical limit with a smoothly varying metric that can be described by the
continuum version of the refractive law.

The paper is organized as follows. In section \ref{sec:C} we recapitulate the classical brachistochrone problem. Although this material can be found at
various places it will be convenient for the reader to present a concise account suited for the present purpose.
In section \ref{sec:M} we defined a non-Euclidean metric of the half-plane such that the brachistochrones are exactly the geodesics of this metric.
The classical brachistochrone problem can be generalized to a situation with a two-dimensional potential having a one-parameter family of symmetries, see
section \ref{sec:G}. Here the Snellius' refraction law acquires an extra factor that compensates the change of the normal direction
needed to define the angle of refraction $\theta$. As an example of this generalization we calculate the brachistochrones for the
harmonic oscillator potential in section \ref{sec:HO}.

Section \ref{sec:R} contains the derivation of the Snellius' law for time-like geodesics for the special case where the metric
has the form of a Schwarzschild (or slightly more general) metric restricted to $(1+1)$ dimensions, namely
\begin{equation}\label{I2}
 d\tau^2 = \varphi(r)\,dt^2-\varphi(r)^{-1}\,\frac{dr^2}{c^2}
 \;,
\end{equation}
$c$ denoting the velocity of light in vacuo.
Note that the Reissner-Nordstr\"om metric describing a charged black hole, see, e.~g.~\cite{W84}, is also of the form (\ref{I2}).
We will show that the Snellius' law is equivalent to the energy conservation law for one-dimensional motion in the Schwarzschild metric, see \cite{W84}, (6.3.12).
Hence it does not represent a new result about general-relativistic free fall motion but rather a new interpretation of a well-known law.

%%%%%%%%%%%%%%%%%%%%%%%%%%%%%%%%%%%%%%%%%%%%%%%%%%%%%%%%%%%%%%%%%%%%%%%%%%%%%%%%%%%%%%%%%%%%%%%%%%%%%%%%%%%%%%%%%%%%%%%%%%%%%%%
\section{The brachistochrone problem revisited}\label{sec:B}
%%%%%%%%%%%%%%%%%%%%%%%%%%%%%%%%%%%%%%%%%%%%%%%%%%%%%%%%%%%%%%%%%%%%%%%%%%%%%%%%%%%%%%%%%%%%%%%%%%%%%%%%%%%%%%%%%%%%%%%%%%%%%%%

%%%%%%%%%%%%%%%%%%%%%%%%%%%%%%%%%%%%%%%%%%%%%%%%%%%%%%%%%%%%%%%%%%%%%%%%%%%%%%%%%%%%%%%%%%%%%%%%%%%%%%%%%%%%%%%%%%%%%%%%%%%%%%%
\subsection{The classical brachistochrone problem}\label{sec:C}
%%%%%%%%%%%%%%%%%%%%%%%%%%%%%%%%%%%%%%%%%%%%%%%%%%%%%%%%%%%%%%%%%%%%%%%%%%%%%%%%%%%%%%%%%%%%%%%%%%%%%%%%%%%%%%%%%%%%%%%%%%%%%%%

\begin{figure}[h]
  \centering
    \includegraphics[width=0.8\linewidth]{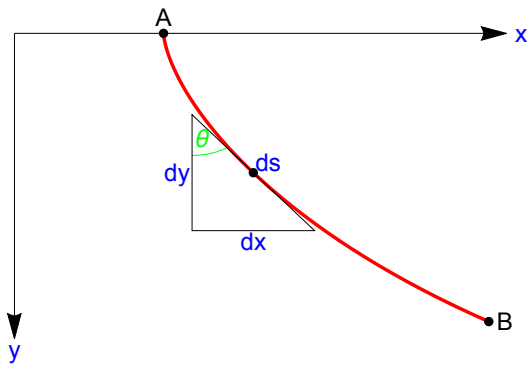}
  \caption[1]
  {A curve  ${\mathcal C}(A,B)$ joining the points $A$ and $B$ and further notations explained in the text.
  }
  \label{FIGSMS1}
\end{figure}

Despite its age of over $300$ years the brachistochrone problem has also found recent attention \cite{HK95}, \cite{B14}, especially in connection
with its generalization including friction \cite{AB75} -- \cite{H05} that was already considered in \cite{E44}.
For the purpose of this paper we will shortly recapitulate its formulation and solution.

Let ${\mathcal C}(A,B)$ be a plane curve between the points $A$ and $B$ and consider the constrained motion of a point particlealong ${\mathcal C}(A,B)$ under the influence of a constant (gravitational) acceleration of absolute value $g$. The initial velocity
at the point $A$ is assumed to vanish. Upon
introducing coordinates $x,y\ge 0$ into horizontal and ``downward" vertical direction, see Figure \ref{FIGSMS1}, the curve
${\mathcal C}(A,B)$ will, at least locally, be described by a smooth function $x\mapsto y(x)$ with derivative
$y'(x)=\frac{dy}{dx}$. Due to the energy conservation
\begin{equation}\label{B1}
  \frac{m}{2}v^2=m\,g\,y
  \;,
\end{equation}
where $m$ is the (irrelevant) mass of the particle, the absolute velocity $v$ of the particle is given by
\begin{equation}\label{B2}
  v=\sqrt{2\,g\,y}
  \;.
\end{equation}
The time $dt$ needed to pass an infinitesimal part of the curve with length $ds$  can be written as
\begin{equation}\label{B3}
 dt= \frac{ds}{v} = \frac{\sqrt{dx^2+dy^2}}{v} \stackrel{(\ref{B2})}{=} \sqrt{\frac{1+y'^2}{2\,g\,y}}\,dx
 \;.
\end{equation}
Hence the total time $T$ to pass the curve ${\mathcal C}(A,B)$ will be given by
\begin{equation}\label{B4}
 T=\int_{x_A}^{x_B}\sqrt{\frac{1+y'(x)^2}{2\,g\,y(x)}}\,dx \;,
\end{equation}
where $(x_A,y_A=0)$ and $(x_B,y_B)$ are the coordinates of the points $A$ and $B$, resp.~.

The brachistochrone problem consists in finding the curve ${\mathcal C}(A,B)$ that makes $T$ minimal for given $A$ and $B$.
We will first recapitulate the approach due to the variational calculus. To this end we re-write (\ref{B4}) in the form
\begin{equation}\label{B5}
 T=\int_{x_A}^{x_B}{\mathcal L}(y(x),y'(x))\,dx \;,
\end{equation}
introducing the Lagrangian
\begin{equation}  \label{B6}
{\mathcal L}(y,y')\equiv \sqrt{\frac{1+y'^2}{2\,g\,y}}
\;.
\end{equation}
Then the solution of the above variational problem is given by a solution of the Euler-Lagrangian equation
\begin{equation}\label{B7}
0=\frac{d}{dx} \frac{\partial {\mathcal L}}{\partial y'}-\frac{\partial {\mathcal L}}{\partial y}
\;.
\end{equation}
Conversely, each solution of the Euler-Lagrange equation yields a solution where the quantity (\ref{B5}) has  locally a stationary value.
We will not consider this equation directly but use the fact that $x$ is a cyclic coordinate of (\ref{B6}) and hence the ``Hamiltonian"
\begin{equation}\label{B8}
  {\mathcal H}\equiv y'\,\frac{\partial {\mathcal L}}{\partial y'}-{\mathcal L}
\end{equation}
is a constant of motion, invoking Noether's theorem.
After evaluating (\ref{B8}) we thus obtain
\begin{equation}\label{B9}
 -{\mathcal H}=\frac{1}{\sqrt{2\,g\,y\,\left(1+y'^2 \right)}}=\frac{\sin \theta}{v}=\mbox{const.}
 \;,
\end{equation}
where the angle $\theta$ is introduced according to Figure \ref{FIGSMS1} and satisfies
\begin{equation}\label{B9a}
  \sin\theta=\frac{dx}{ds}=\frac{dx}{\sqrt{dx^2+dy^2}}=\frac{1}{\sqrt{1+y'^2}}
  \;.
\end{equation}
Formulated in this way the conservation law (\ref{B9})
assumes the form of Snellius' law of refraction. Moreover, the latter is equivalent to
\begin{equation}\label{B10}
  2\,R\equiv y\,\left( 1+y'^2\right) =\mbox{const.}
  \;,
\end{equation}
where $R\ge 0$ is some parameter. Strictly speaking, this equivalence only holds if the case ${\mathcal H}=0$
that corresponds to the limit $R\rightarrow\infty$ is treated separately.
By contrast, the case of $R=0$ has to be excluded since it leads to $y(x)\equiv 0$ and hence the integral (\ref{B4}) would diverge.
The remaining derivations will only assume (\ref{B10}) and not the Euler--Lagrange equation (\ref{B7}).

\begin{figure}[h]
  \centering
    \includegraphics[width=0.8\linewidth]{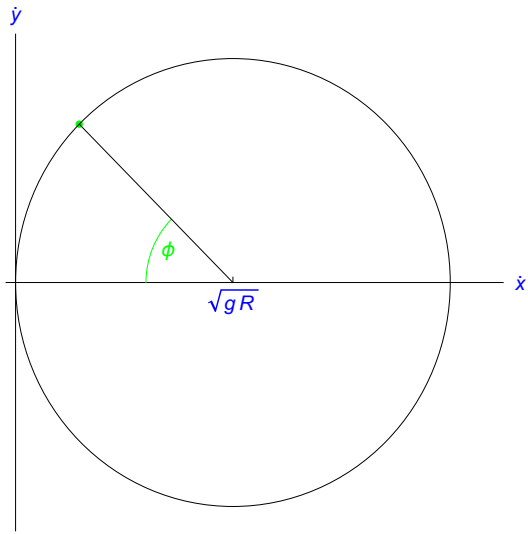}
  \caption[2]
  {The circle in the $\dot{x}-\dot{y}-$plane with parameter representation (\ref{B13a}),(\ref{B13b}) .
  }
  \label{FIGSMS2}
\end{figure}

(\ref{B10}) has the form of an implicit differential equation of first order and can be solved by separation of variables.
However, this would require some awkward calculations and hence we will proceed in a different way.
We will rather write down the time derivatives of $x$ and $y$, denoted by a dot,
where the time dependence of the involved functions is usually suppressed:
\begin{eqnarray}
\label{B11a}
\dot{x}  &\stackrel{(\ref{B3})}{=}& \sqrt{\frac{2 g y}{1+y'^2}}\stackrel{(\ref{B10})}{=} \sqrt{\frac{g}{R}}\,y,\\
\label{B11b}
 \dot{y} &=& y'\,\dot{x}\stackrel{(\ref{B11a})}{=} \sqrt{\frac{g}{R}}\,y\,y'
 \stackrel{(\ref{B10})}{=} \sqrt{\frac{g}{R}}\,\sqrt{y(2R-y)}
 \;.
\end{eqnarray}
We will shortly comment on the sign ambiguity introduced by the square root in (\ref{B11b}). The positive sign chosen in (\ref{B11b})
holds for the descending part of the brachistochrone. It will be tacitly understood in what follows that there exists also
an ascending part where a negative sign would have to be inserted into (\ref{B11b}).

Since both derivatives $\dot{x}$ and $\dot{y}$ only depend on $y$ this can be viewed as a parameter representation of a curve
in the $\dot{x}-\dot{y}-$plane. The form of the curve follows from the following calculation:
\begin{eqnarray}
\label{B12a}
  \dot{y}^2 &\stackrel{(\ref{B11b})}{=}& \frac{g}{R}\,y(2R-y) \\
  \label{B12b}
  &=& \frac{g}{R}\,\left(R^2-(y-R)^2)\right)\\
  \label{B12c}
  &\stackrel{(\ref{B11a})}{=}& \frac{g}{R}\,\left(R^2-\left( \sqrt{\frac{R}{g}}\,\dot{x}-R\right)^2\right)\\
  \label{B12d}
  &=& \frac{g}{R}\,\left(R^2-\frac{R}{g}\left( \dot{x}-\sqrt{g\,R}\right)^2\right)\\
  \label{B12e}
  &=& g\,R -\left(\dot{x}-\sqrt{g\,R} \right)^2
  \;.
 \end{eqnarray}
 It is a circle with center $(\sqrt{g\,R},0)$ and radius $\sqrt{g\,R}$, see Figure \ref{FIGSMS2}.
 Another parameter representation of this circle uses the angle $\phi$, see Figure \ref{FIGSMS2},
 \begin{eqnarray}
 \label{B13a}
  \dot{x}  &=& \sqrt{g\,R}\,  \left( 1-\cos\phi\right), \\
  \label{B13b}
  \dot{y} &=& \sqrt{g\,R}\,\sin\phi
  \;.
 \end{eqnarray}
 Differentiating (\ref{B13a}) w.~r.~t.~time yields
 \begin{equation}\label{B14}
   \ddot{x}=\sqrt{g\,R}\,\sin\phi\;\dot{\phi}
   \;.
 \end{equation}
On the other hand, we may differentiate (\ref{B11a}) and obtain
\begin{equation}\label{B15}
   \ddot{x}=\sqrt{\frac{g}{R}}\,\dot{y}\stackrel{(\ref{B13b})}{=}\sqrt{\frac{g}{R}}\,\sqrt{g\,R}\,\sin\phi
   \;.
 \end{equation}
 Comparison of (\ref{B14}) and (\ref{B15}) shows that the angular velocity $\dot{\phi}$ assumes the constant value
 \begin{equation}\label{B16}
  \dot{\phi}=\sqrt{\frac{g}{R}}
  \;.
 \end{equation}
 Since both $t=0$ and $\phi=0$ at the point $A$ we may further conclude
 \begin{equation}\label{B17}
   \phi(t)=\sqrt{\frac{g}{R}}\,t
   \;.
 \end{equation}
 This enables the $t$-integration of (\ref{B13a}) and (\ref{B13b}) in a straightforward manner with the result
 \begin{eqnarray}
 \label{B18a}
   x(t) &=& x_A+\sqrt{g R}\,t-\sqrt{g R}\sqrt{\frac{R}{g}}\sin \sqrt{\frac{g}{R}}t \\
   \label{B18b}
    &=& x_A+R\,\left(\phi(t)-\sin\phi(t)\right),\\
    \label{B18c}
   y(t)&\stackrel{(\ref{B11a})}{=}& \sqrt{\frac{R}{g}}\,\dot{x}\\
   \label{B18d}
   &\stackrel{(\ref{B13a})}{=}& R\,\left(1-\cos\phi(t)\right)
   \;.
 \end{eqnarray}
 This is obviously the parameter representation of a cycloid,
the curve traced by a point on the rim of a circular wheel with radius $R$ as the wheel rolls along a straight line without slipping.
This completes the solution of the classical brachistochrone problem.

\begin{figure}[h]
  \centering
    \includegraphics[width=0.8\linewidth]{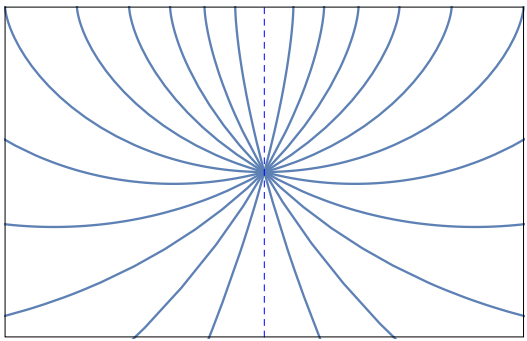}
  \caption[3]
  {A family of cycloids, geodesics of the metric (\ref{B19}), passing through a point of the half-plane ${\sf H}$.
  The limit case $R\rightarrow\infty$ is indicated by a vertical dashed line.
  }
  \label{FIGSMS3}
\end{figure}

%%%%%%%%%%%%%%%%%%%%%%%%%%%%%%%%%%%%%%%%%%%%%%%%%%%%%%%%%%%%%%%%%%%%%%%%%%%%%%%%%%%%%%%%%%%%%%%%%%%%%%%%%%%%%%%%%%%%%%%%%%%%%%%
\subsection{The classical brachistochrone metric}\label{sec:M}
%%%%%%%%%%%%%%%%%%%%%%%%%%%%%%%%%%%%%%%%%%%%%%%%%%%%%%%%%%%%%%%%%%%%%%%%%%%%%%%%%%%%%%%%%%%%%%%%%%%%%%%%%%%%%%%%%%%%%%%%%%%%%%%

We may reformulate the brachistochrone problem in a slightly different language by writing the square of (\ref{B3}) as
\begin{equation}\label{B19}
 dt^2 = \frac{dx^2+dy^2}{2\,g\,y}
 \;,
\end{equation}
and viewing this equation as the definition of a Riemannian metric in the open half-plane ${\sf H}$ given by
\begin{equation}\label{B20}
  {\sf H}=\{(x,y)\in {\mathbbm R}^2\,\left| y>0\,\right.\}
  \;.
\end{equation}
This approach is well-known from geometrical optics, see, e.~g.~\cite{B14}.
The length of a curve $ {\mathcal C}(A,B)$ in ${\sf H}$ w.~r.~t.~this metric is the time that a point particle constrained to
move on ${\mathcal C}(A,B)$ needs to run through the curve, similarly as described above. The only difference is that the starting point
$A$ of the curve can be chosen arbitrarily in ${\sf H}$ and hence
the particle has a non-vanishing initial velocity $v_A=\sqrt{2 \,g\, y_A}$ according to (\ref{B2}). Then the preceding
considerations show that the family of cycloids defined by (\ref{B18b}) and (\ref{B18d}), where the two parameters $x_A\in{\mathbbm R}$ and $R>0$
run through all allowed values, is exactly the family of all geodesics of the metric (\ref{B19}).
To be a bit more precise, we have to extend the set of cycloids defined above by straight vertical lines,
that could be viewed as the limit of the curves defined by (\ref{B18b}) and (\ref{B18d}) for $R\rightarrow \infty$. Then it follows that
for each point $A\in{\sf H}$ and each direction ${\mathbf t}$ there exists a cycloid (in the extended sense) that passes through $A$
and is tangent to ${\mathbf t}$, see Figure \ref{FIGSMS3}.
It remains an open problem to further analyze the metric (\ref{B19}) and to decide whether it is isometric to a known structure.
As a first result into this direction we mention that the scalar curvature ${\sf R}$ of the metric (\ref{B19}) is given by
\begin{equation}\label{B21}
  {\sf R}=-\frac{4 \,g}{y}
  \;,
\end{equation}
and hence negative and not constant.

%%%%%%%%%%%%%%%%%%%%%%%%%%%%%%%%%%%%%%%%%%%%%%%%%%%%%%%%%%%%%%%%%%%%%%%%%%%%%%%%%%%%%%%%%%%%%%%%%%%%%%%%%%%%%%%%%%%%%%%%%%%%%%%
\subsection{The generalized brachistochrone problem}\label{sec:G}
%%%%%%%%%%%%%%%%%%%%%%%%%%%%%%%%%%%%%%%%%%%%%%%%%%%%%%%%%%%%%%%%%%%%%%%%%%%%%%%%%%%%%%%%%%%%%%%%%%%%%%%%%%%%%%%%%%%%%%%%%%%%%%%

We may generalize the classical brachistochrone problem by considering a more general two-dimensional potential and (local) orthogonal coordinates
$x,y$ such that the analogue of (\ref{B1}) leads to a velocity field $v(y)$ and the Euclidean metric has the form
\begin{equation}\label{G1}
  ds^2=g_1(y)\,dx^2+g_2(y)\,dy^2
  \;.
\end{equation}
It is thus invariant under the one-parameter group of translations into $x-$ direction.
Then the analogue variational problem leads to a Lagrangian
\begin{equation}\label{G2}
{\mathcal L}(y,y')=\frac{\sqrt{g_1(y)+g_2(y)\,y'^2}}{v(y)}
\;.
\end{equation}
By assumption, $x$ is still a cyclic coordinate and hence the ``Hamiltonian"
\begin{equation}\label{G3}
 {\mathcal H}=y'\frac{\partial {\mathcal L}}{\partial y'}-{\mathcal L}=-\frac{g_1}{v\,\sqrt{g_1+g_2\,y'^2}}
\end{equation}
will be a constant of motion again invoking Noether's theorem. The angle $\theta$ between the brachistochrone and the local $y$-direction now satisfies
\begin{equation}\label{G4}
  \sin\theta =\sqrt{\frac{g_1}{g_1+g_2 \,y'^2}}
\end{equation}
and hence the Snellius law of refraction assumes the form
\begin{equation}\label{G5}
  -{\mathcal H}=\frac{\sqrt{g_1}\sin\theta}{v}=C=\mbox{const.}
\;.
\end{equation}
After a straightforward calculation the analogue of (\ref{B11a}) and (\ref{B11b}) is obtained as
\begin{eqnarray}
\label{G6a}
  \dot{x} &=& \frac{v^2}{g_1}C, \\
  \label{G6b}
  \dot{y}&=& -\frac{v}{\sqrt{g_1\,g_2}}\,\sqrt{g_1-v^2\,C^2}
  \;.
\end{eqnarray}
This describes again a curve in the $\dot{x}-\dot{y}-$ plane but its form depends on the potential and an explicit solution
analogous to the classical brachistochrone problem is not possible in general.

\begin{figure}[h]
  \centering
    \includegraphics[width=0.8\linewidth]{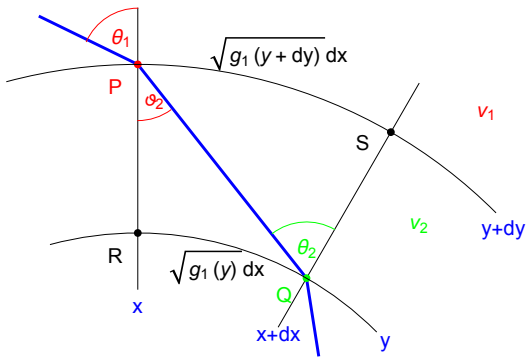}
  \caption[6]
  {The geometry of the refraction law for the generalized brachistochrone problem. Since the direction normal to the
  equipotential lines $y=\mbox{const.}$ changes from $P$ to $Q$ the correction factor $\sqrt{g_1}$ occurs
  in (\ref{G5}) and (\ref{G9}).
  }
  \label{FIGSMS6}
\end{figure}

It remains to make the factor$ \sqrt{g_1}$ in the refraction law (\ref{G5}) plausible. To this end we consider an
infinitesimal quadrangle $P,R,Q,S$, see Figure \ref{FIGSMS6}, and the light-ray from $P$ to $Q$. The Snellius' law only
yields a relation between $\theta_1$ and the alternate angle $\vartheta_2$, namely
\begin{equation}\label{G7}
\frac{\sin\theta_1}{v_1}=\frac{\sin\vartheta_2}{v_2}
\;.
\end{equation}
We cannot assume that $\vartheta_2=\theta_2$ holds in general since the normal to the equipotential lines $y=\mbox{const.}$ has
different directions at $P$ and $Q$. This difference is of first order in the distance $\overline{P\,Q}$ and hence of the
same order as the change of the velocity $v$. More precisely, in the first order we obtain
\begin{eqnarray}
\label{G8a}
 \sin\vartheta_2 &=& \frac{\overline{R\,Q}}{\overline{P\,Q}} =\frac{\sqrt{g_1(y)}dx}{\overline{P\,Q}},\\
 \label{G8b}
 \sin\theta_2 &=& \frac{\overline{P\,S}}{\overline{P\,Q}} =\frac{\sqrt{g_1(y+dy)}dx}{\overline{P\,Q}}.
\end{eqnarray}
Together with (\ref{G7}) it follows that
\begin{equation}\label{G9}
  \frac{\sqrt{g_1(y+dy)}\,\sin\theta_1}{v_1}=\frac{\sqrt{g_1(y)}\,\sin\theta_2}{v_2}
  \;,
\end{equation}
and hence (\ref{G5}) holds.

%%%%%%%%%%%%%%%%%%%%%%%%%%%%%%%%%%%%%%%%%%%%%%%%%%%%%%%%%%%%%%%%%%%%%%%%%%%%%%%%%%%%%%%%%%%%%%%%%%%%%%%%%%%%%%%%%%%%%%%%%%%%%%%
\subsection{Example: The harmonic oscillator brachistochrone problem}\label{sec:HO}
%%%%%%%%%%%%%%%%%%%%%%%%%%%%%%%%%%%%%%%%%%%%%%%%%%%%%%%%%%%%%%%%%%%%%%%%%%%%%%%%%%%%%%%%%%%%%%%%%%%%%%%%%%%%%%%%%%%%%%%%%%%%%%%

As an example of the generalized brachistochrone problem we consider the two-dimensional harmonic oscillator potential
\begin{equation}\label{HO1}
 V(\rho) =\frac{m}{2}\,\omega^2\,\rho^2
  \;.
\end{equation}
where $\rho,\phi$ are polar coordinates such that the Euclidean metric assumes the form
\begin{equation}\label{HO2}
  ds^2=\rho^2\,d\phi^2+d\rho^2
  \;.
\end{equation}
We choose the coordinates $x=\phi$ and $y=\sigma\equiv \rho^2$ such that
\begin{equation}\label{HO3}
  ds^2=g_1(\sigma)\,d\phi^2+g_2(\sigma)\,d\sigma^2=\sigma\,d\phi^2+\frac{1}{4\sigma}\,d\sigma^2
   \;.
\end{equation}
Let the starting point $A$ of the brachistochrone have the coordinate $\rho_A=R$ such that the velocity field can be written as
\begin{equation}\label{HO4}
v(\sigma)=\omega \sqrt{R^2-\sigma}
\;,
\end{equation}
and the refraction law reads
\begin{equation}\label{HO5}
 \frac{\sqrt{\sigma}\,\sin\theta}{v}=C=\mbox{const.}
 \;.
\end{equation}

\begin{figure}[h]
  \centering
    \includegraphics[width=0.8\linewidth]{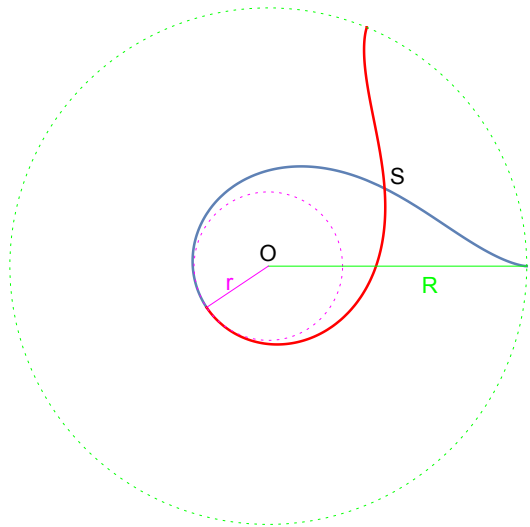}
  \caption[5]
  {The  brachistochrone resulting from a two-dimensional harmonic oscillator potential, consisting of a descending part (blue curve) and an ascending part (red curve).
  Both parts cross at the point $S$.
  The chosen values are $\omega=1\,s^{-1},\,R=4\,m$   and $C=0.3\,s$.  It follows from (\ref{HO9}) that the minimal radius $r$ is reached after the time $T=1.50455\,s$.
  To pass the straight (green) line of length $R$ from the starting point to the center $O$
  would already require the longer time $T_1=\frac{\pi}{2\,\omega}\approx 1.5708\,s$. To pass the closed part of the brachistochrone from $S$ to $S$ requires
  $T_2=0.956653\, s$, whereas one would need $1.14388\,s$ to go from $S$ on a straight line to the center $O$ and back.
  }
  \label{FIGSMS5}
\end{figure}

Then (\ref{G6a}) and (\ref{G6b}) assume the form
\begin{eqnarray}
\label{HO6a}
  \dot{\phi} &=& \frac{\omega^2\left( R^2-\sigma\right)}{\sigma} \,C,\\
 \label{HO6b}
  \dot{\sigma} &=& -2\,\omega\,\sqrt{1+C^2\,\omega ^2}\,\sqrt{\left(R^2-\sigma \right) \left(\sigma -r^2\right)},
    \end{eqnarray}
  where
  \begin{equation}\label{HO6c}
  r\equiv \frac{C\omega  R}{\sqrt{1+C^2 \omega ^2}}
   \end{equation}
is the minimal radius of the brachistochrone, see Figure \ref{FIGSMS5}.

We may solve the differential equation (\ref{HO6b}) by separation of variables, insert the result into (\ref{HO6a}) and finally integrate over $t$.
The result allowing for the initial condition $\sigma(0)=R^2$  reads
\begin{equation}\label{HO7}
\sigma(t)=  \frac{R^2 \left(\cos \left(2  \omega t \sqrt{1+C^2 \omega ^2}\right)+2 C^2 \omega ^2+1\right)}
  {2   \left(1+C^2 \omega ^2\right)}
  \;,
\end{equation}
and
\begin{equation}\label{HO8}
  \phi(t)=
  \phi _0- \omega ^2 t+\frac{1}{C}\,\arctan \left(\frac{C \omega  \tan \left( \omega \,t\, \sqrt{1+C^2 \omega
   ^2}\right)}{\sqrt{1+C^2 \omega ^2}}\right).
\end{equation}
It follows that the minimal radius $r$ is reached after the time
\begin{equation}\label{HO9}
 T= \frac{\pi }{2 \omega  \sqrt{1+C^2 \omega ^2}}
 \;.
\end{equation}

An example of the resulting brachistochrone is shown in Figure \ref{FIGSMS5}. Here we see that the brachistochrone of the harmonic oscillator
potential will have a point $S$ of self-intersection. The minimal time to pass from $S$ to $S$ is, of course, $T=0$, whereas one would need
a finite time $T_2$  to follow the brachistochrone from $S$ to $S$ by winding once around the origin $O$.
This is no contradiction since the brachistochrone only represents
a {\it local} minimum of time, not a global one, as it is illustrated by numerical examples in Figure \ref{FIGSMS5}.

%%%%%%%%%%%%%%%%%%%%%%%%%%%%%%%%%%%%%%%%%%%%%%%%%%%%%%%%%%%%%%%%%%%%%%%%%%%%%%%%%%%%%%%%%%%%%%%%%%%%%%%%%%%%%%%%%%%%%%%%%%%%%%
\section{Refraction of time-like geodesics}\label{sec:R}
%%%%%%%%%%%%%%%%%%%%%%%%%%%%%%%%%%%%%%%%%%%%%%%%%%%%%%%%%%%%%%%%%%%%%%%%%%%%%%%%%%%%%%%%%%%%%%%%%%%%%%%%%%%%%%%%%%%%%%%%%%%%%%%

As pointed out in the Introduction we will look for an analogue of Snellius' law in the theory of freely falling particles
according to General Relativity (GR). It will be advisable to stress the differences to the classical brachistochrone problem (BP) in order to avoid
misunderstandings. Both problems have to do with the motion of point particles in a gravitational field but the GR case deals with {\it free}
fall in contrast to the motion constrained to a curve ${\mathcal C}(A,B)$ in the BP. Here $A$ and $B$ are two points
in the (closure of the) half-space ${\sf H}$ defined in Section \ref{sec:B} and the brachistochrone realizes the minimal (non-relativistic) time.
The world-line of a freely falling particle in GR realizes
the maximal {\it proper time} among all time-like curves that connect two points $A$ and $B$ in {\it spacetime}.
Light rays or paths only occur in the BP by means of analogy. They do not play any role in the considered GR case although Fermat's principle
also holds in GR, see \cite{P00}, theorem 7.3.1.
But despite these differences both solution curves are geodesics of a (pseudo) Riemannian metric and hence solutions of an Euler-Lagrange equation.
This common ground means that they can be investigated by similar techniques.

\begin{figure}[h]
  \centering
    \includegraphics[width=0.8\linewidth]{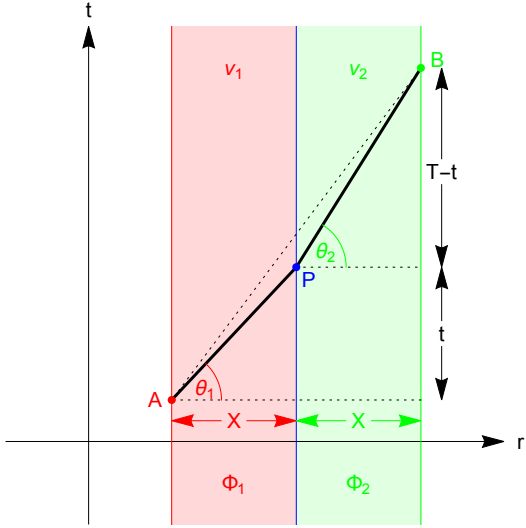}
  \caption[4]
  {The geometry of the Snellius' law for two regions with different velocities $v_1>v_2$. The same figure can  be used to illustrate
  the spacetime geometry of the corresponding law for two spacetime regions with different gravitational potentials (values of the metric)
  $\Phi_1 < \Phi_2$.
  }
  \label{FIGSMS4}
\end{figure}

It will be instructive to repeat the well-known derivation of the Snellius' law by using Fermat's principle. For the illustration we will use the
same Figure as for the derivation of the refraction law for time-like geodesics and hence adopt an apparently strange notation
denoting the vertical coordinate by $t$, the horizontal one by $r$ and the time by $\tau$.
We assume that the velocity of light has two different but constant values $v_1$ and $v_2$ in two regions ${\mathcal A}$ and ${\mathcal B}$
separated by plane (the blue line in Figure \ref{FIGSMS4}). Hence according to Fermat's principle the light will choose
straight lines inside the regions ${\mathcal A}$ and ${\mathcal B}$. Thus the total path from $A\in {\mathcal A}$ to
$B\in {\mathcal B}$ will be composed of a straight line
joining $A$ and $P$, and another straight line joining $P$ and $B$
where $P$ is an arbitrary point at the boundary between ${\mathcal A}$ and ${\mathcal B}$.
The total time to travel from $A$ to $B$ then amounts to
\begin{equation}\label{R1}
 \tau=\tau_1+\tau_2 = \frac{\sqrt{t^2+X^2}}{v_1}+\frac{\sqrt{(T-t)^2+X^2}}{v_2}
 \;,
\end{equation}
cf.~the notation introduced in Figure \ref{FIGSMS4}.
It will be a minimum for
\begin{equation}\label{R2}
0=\frac{\partial \tau}{\partial t}=\frac{t}{v_1\,\sqrt{t^2+X^2}}-\frac{T-t}{v_2\,\sqrt{(T-t)^2+X^2}}
\;,
\end{equation}
or, equivalently,
\begin{equation}\label{R3}
  \frac{\sin\theta_1}{v_1}= \frac{\sin\theta_2}{v_2}
  \;,
\end{equation}
where the angles $\theta_1$ and $\theta_2$ are defined according to Figure \ref{FIGSMS4}.
This is Snellius' law of refraction;
the version that covers also the case of a continuously varying velocity field reads
\begin{equation}\label{R4}
  \frac{\sin\theta}{v}=C=\mbox{const.}
\end{equation}
and appeared in the above solution of the classical brachistochrone problem in equation (\ref{B9}).
In the continuous case the light path cannot enter regions with $|C\,v|>1$ and has a vertical tangent
(w.~r.~t.~Figure \ref{FIGSMS4}) at places where $|C\,v|=1$ due to total internal reflection.

Next we will perform the analogous calculation for time-like curves in two-dimensional spacetime with the metric (\ref{I2}).
We will assume that the function $\phi(r)$ that occurs in (\ref{I2}) has two different but constant values $\Phi_1>0$ and $\Phi_2>0$
in the spacetime regions ${\mathcal A}$ and ${\mathcal B}$, resp., and calculate the total proper time $\tau$ to travel from
$A$ to $B$ via $P$:
\begin{equation}\label{R5}
\tau=\tau_1+\tau_2= \sqrt{\Phi_1\,t^2-\frac{X^2}{\Phi_1\,c^2}}+\sqrt{\Phi_2\,(T-t)^2-\frac{X^2}{\Phi_2\,c^2}}
 \;.
\end{equation}
It will be a maximum for
\begin{eqnarray}
\label{R6a}
  0 &=& \frac{\partial \tau}{\partial t}\\
  \label{R6b}
   &=& \frac{\Phi_1\,t}{\sqrt{\Phi_1\,t^2-\frac{X^2}{\Phi_1\,c^2}}}\\
   \label{R6c}
  &&- \frac{\Phi_2\,(T-t)}{\sqrt{\Phi_2\,(T-t)^2-\frac{X^2}{\Phi_2\,c^2}}}
  \;.
\end{eqnarray}
In order to cast this equation into a similar form as (\ref{R3}) we introduce positive variables $a,b,u_1,u_2$ such that
\begin{eqnarray}
\label{R7a}
  \sqrt{\Phi_1}\,t &=& a\,\cosh u_1, \\
\label{R7b}
  \frac{X}{\sqrt{\Phi_1}\,c}&=& a\,\sinh u_1, \\
\label{R7c}
 \sqrt{\Phi_2}\,(T-t) &=& b\,\cosh u_2, \\
\label{R7d}
  \frac{X}{\sqrt{\Phi_2}\,c}&=& b\,\sinh u_2
  \;.
\end{eqnarray}
This implies that
\begin{equation}\label{R8}
  \tanh\,u_1=\frac{X}{c\,t\,\Phi_1}
\end{equation}
is the velocity of the particle in the region ${\mathcal A}$,
measured w.~r.~t.~the metric in this region using the constant value $\Phi_1$,
analogously for
\begin{equation}\label{R9}
  \tanh\,u_2=\frac{X}{c\,(T-t)\,\Phi_2}
  \;.
\end{equation}
Then the ``refraction law" (\ref{R6a}) - (\ref{R6c}) can be concisely written as
\begin{equation}\label{R10}
  \sqrt{\Phi_1}\,\cosh\,u_1= \sqrt{\Phi_2}\,\cosh\,u_2
  \;.
\end{equation}

The version that also covers the case of a continuously varying metric reads
\begin{equation}\label{R11}
  \sqrt{\varphi(r)}\,\cosh\,u=E=\mbox{const.}
  \;,
\end{equation}
where $u$ is implicitly defined by the analogue of (\ref{R8}):
\begin{equation}\label{R12}
 \tanh\,u= \frac{dr}{c\,dt\,\varphi}
 \;.
\end{equation}
In the continuous case the world-line cannot enter regions with $\frac{E}{ \sqrt{\varphi(r)}}<1$ and has a vertical tangent
(w.~r.~t.~Figure \ref{FIGSMS4}) and hence a vanishing velocity at places where $\frac{E}{ \sqrt{\varphi(r)}}=1$
due to the analogue of total internal reflection.

An equivalent form of the continuous refraction law is obtained by the following calculation
\begin{eqnarray}
\label{R13a}
  d\tau &=& \sqrt{\varphi\,dt^2-\frac{dr^2}{\varphi\,c^2}} \\
   &=& \sqrt{\varphi}\,dt\,\sqrt{1-\left(\frac{dr}{c\,\varphi\, dt}\right)^2}\\
   &\stackrel{(\ref{R12})}{=}& \sqrt{\varphi}\,dt\,\sqrt{1-\tanh^2\,u}  \\
   &=& \frac{\sqrt{\varphi}\,dt}{\cosh\,u}
   \;,
\end{eqnarray}
and hence
\begin{equation}\label{R14}
  \varphi\,\frac{dt}{d\tau}=\sqrt{\varphi}\cosh\, u=E=\mbox{const.}
  \;.
\end{equation}

We remark that (\ref{R14}) is not a new equation but well-known. Consider the case of the Schwarzschild metric, i.~e.~, (\ref{I2}) with the special choice
 \begin{equation}\label{R15}
  \varphi(r)=1-\frac{2\, G\, M}{c^2\,r}
  \;,
 \end{equation}
where, as usual, $G$ denotes the gravitational constant and $M$ the mass of the gravitational center.
Then equation (\ref{R14}) is equivalent to (6.3.12) in \cite{W84} and the latter  follows from the fact that the Schwarzschild metric is static.
The quantity (\ref{R14}) hence represents some sort of energy. This is reminiscent to the brachistochrone problem where the Snellius law
can be derived from the fact that the ``Hamiltonian" ${\mathcal H}$ will be constant of motion, see Section \ref{sec:B}.

Finally, we will consider terrestrial free fall experiments with a height of, say, $y\approx 1\,m$
and estimate the order of magnitude of the involved quantities.
Let $r_0=6.371\times 10^6\,m$ be the radius of the earth such that
\begin{equation}\label{R16}
 \frac{M\,G}{r_0^2}=g=9.81 \frac{m}{s^2}
 \;.
\end{equation}
Then we  write $r=r_0+y$ and expand  $\sqrt{\varphi(r)}$ according to the Schwarzschild metric w.~r.~t.~$y$ in the following way:
\begin{equation}\label{R17}
 \sqrt{\varphi(r)}=\sqrt{1-\frac{2\,G\,M}{c^2(r_0+y)}}\stackrel{(\ref{R16})}{\approx} 1-\frac{r_0\,g}{c^2}+\frac{g\,y}{c^2}
 \;,
\end{equation}
Since the velocities $v$ involved are small compared with $c$  we may set $u\approx \frac{v}{c}$ and hence
\begin{equation}\label{R18}
 \cosh\,u \approx 1+\frac{v^2}{2\,c^2}
 \;.
\end{equation}
Then the law of refraction (\ref{R11}) leads to
\begin{equation}\label{R19}
  \sqrt{\varphi(r)}\,\cosh\,u \stackrel{(\ref{R17})(\ref{R18})}{\approx} 1-\frac{r_0\,g}{c^2}+\frac{\frac{1}{2}v^2+g\,y}{c^2}=\mbox{const.}
  \;,
\end{equation}
thus recovering the non-relativistic energy conservation law (\ref{B1}). Although the relative variation
of $\varphi(r)$ is of the order of $10^{-16}$ this is compatible with the change of the non-relativistic quantities $\frac{1}{2}v^2$ and $g\,y$
of the order of a few $\frac{m^2}{s^2}$ since these quantities are divided by $c^2$ in (\ref{R19}).

%%%%%%%%%%%%%%%%%%%%%%%%%%%%%%%%%%%%%%%%%%%%%%%%%%%%%%%%%%%%%%%%%%%%%%%%%%%%%%%%%%%%%%%%%%%%%%%%%%%%%%%%%%%%%%%%%%%%%%%%%%%%%%%%%%%%%%%%%%
\section*{Acknowledgment}
%%%%%%%%%%%%%%%%%%%%%%%%%%%%%%%%%%%%%%%%%%%%%%%%%%%%%%%%%%%%%%%%%%%%%%%%%%%%%%%%%%%%%%%%%%%%%%%%%%%%%%%%%%%%%%%%%%%%%%%%%%%%%%%%%%%%%%%%%
I gratefully acknowledge discussions with Thomas Br\"ocker on the subject of this paper.

\end{document}